\documentclass[12pt]{article}
{\begin{Sbox}\begin{minipage}}%
{\end{minipage}\end{Sbox}\fbox{\TheSbox}}%
\usepackage{booktabs}
\usepackage{graphicx}
\usepackage{float}
\usepackage{subfigure}
\usepackage{sidecap}
\usepackage[psamsfonts]{amssymb}
\usepackage{amsmath, amsthm, anysize, enumerate, color, url}
\usepackage{epstopdf}%%\makeatletter \renewcommand\@biblabel[8]{${#8}$}\makeatother
\usepackage{longtable}
\usepackage{bm}
\usepackage[colorlinks,citecolor=blue,urlcolor=blue]{hyperref}
\usepackage{amsmath}
\usepackage{mathrsfs}
\usepackage{geometry}
\usepackage{caption}
\usepackage{subfigure}
\usepackage{float}
\usepackage{breakurl}

\numberwithin{equation}{section}%
\makeatletter
\newcommand\Hsub{\@startsection{subsection}{2}%
 {0pt}{-\baselineskip}{.2\baselineskip}%
 {\normalsize\upshape}}
\makeatother
\newcommand{\pozhehao}{\kern0.3ex\rule[0.8ex]{1.5em}{0.095ex}\kern0.3ex}

\usepackage{setspace}

\begin{document}

\vspace*{0.15cm}
\begin{center}
{\Large\bf Forecasting  confirmed cases of the COVID-19 pandemic with a {migration-based} epidemiological model}
\vskip 1.2\baselineskip
{\large Xinyu Wang$^{1}$, \ \ Lu Yang$^{1}$, \ \ Hong Zhang$^{1}$, \ \ Zhouwang Yang$^{{1^*}}$, \ \ Catherine Liu$^{{2^*}}$}
\vskip 0.2\baselineskip
{\sl University of Science and Technology of China$^{1}$}
\vskip 0.1\baselineskip
\vskip 0.2\baselineskip
{\sl Department of Applied Mathematics, The Hong Kong Polytechnic University$^{2}$}
\vskip 0.1\baselineskip

\end{center}
\footnotetext[1]{$^*$ and $^{2^*}$Correspondence author.\\
Email addresses: abwxy$@$mail.ustc.edu.cn (X. Wang), yl0501$@$mail.ustc.edu.cn (L. Yang),\\ zhangh$@$ustc.edu.cn (H. Zhang), yangzw$@$ustc.edu.cn (Z. Yang),\\ macliu@polyu.edu.hk (C. Liu)}
%fanyifan${\_}7@$163.com (Y. Fan), zhanghuiming$@$pku.edu.cn (H. Zhang ),\\ tingyanty$@$mail.ccnu.edu.cn (T. Yan)}.
\vskip 1.5mm

\begin{abstract}
The unprecedented coronavirus disease 2019 (COVID-19) pandemic is still a worldwide threat to human life since its invasion {into the} daily lives of the public in the first several  months of 2020. Predicting the size of confirmed cases is important for countries and communities to make proper prevention and control policies so as to effectively curb the spread of COVID-19. Different from the 2003 SARS epidemic and the worldwide 2009 H1N1 influenza pandemic, COVID-19 has unique epidemiological characteristics  in its infectious and recovered compartments. This drives us to formulate a new infectious dynamic model for forecasting the COVID-19 pandemic within the human mobility network, named the SaucIR-model in the sense that the new compartmental model extends the benchmark SIR model by dividing the {flow} of people in the infected state into asymptomatic, pathologically infected but unconfirmed, and confirmed. Furthermore, we employ dynamic modeling of population flow in the model in order that spatial effects can be incorporated effectively. We forecast the spread of accumulated confirmed cases in some provinces of mainland China and other countries that experienced severe infection during the time period from late February to early May 2020. The novelty of incorporating the geographic spread of the pandemic leads to a surprisingly good agreement with published confirmed case reports. The numerical analysis validates the high degree of predictability of our proposed SaucIR model compared to existing {resemblance}. The proposed forecasting SaucIR model is implemented in Python. A web-based application is also developed by Dash (under construction).


{\noindent } 	
\end{abstract}

\vskip 5 pt

\section{Introduction}
The conventional susceptible-infected-recovered (SIR) {dynamic} model is one of the simplest compartmental models in epidemiology that segments the flows of people into three states, i.e., susceptible (S), infected (I), and resistant/recovered (R). The SIR model is used to compute the theoretical number of people infected with a contagious illness in a closed population over time \cite{WOK1927}. It has been widely applied and expanded  to estimate or to predict the spread size of contagion phenomena  such as the worldwide 2009 H1N1 influenza pandemic and severe acute respiratory syndrome (SARS) in 2003 {for the purpose of infection prevention and control and public health strategies \cite{Brockmann1337}}. Modeling dynamics of coronavirus disease 2019, COVID-19 in short, remains a big challenge, although there has been sporadic investigation in the past few months \cite{hao2020reconstruction}. The goal of this project is to to develop an expansion of the SIR model, named SaucIR, in response to the spread of the COVID-19 pandemic on the basis of its specific epidemiological characteristics and dynamic migration.

Unfortunately, the officially deployed interventions based on the SIR model were invalidated at the outbreak of COVID-19, particularly at the early stage when COVID-19 burst onto the global pandemic scene, even though the SARS 2003 epidemic and COVID-19 share many similarities. According to the editorial of New England Journal of Medicine (NEJM) in April 2020, the rapid, worldwide spread of COVID-19 resulted in more than 2.6 million people infected within five months, in contrast to the fact that SARS 2003 was controlled within 8 months with less than ten thousand persons infected in limited geographic areas \cite{NEJMe2009758}. More worldwide dynamic data of {the} infection spread are  available to the public \cite{worlddatawebsite}. There are quite a few key factors to interpret the dramatically different trajectories of  transmission and spread between COVID-19 and SARS 2003. For instance, one epidemiological key influence factor is the existence of asymptomatic individuals, who are the silent carriers of coronavirus. Such asymptomatic infections were diagnosed with positive RT-PCR test results but without any relevant clinical symptoms in the preceding days or during hospitalization, inducing the risk of  spreading the disease and hence preventing ascertainment before symptoms \cite{NEJMe2009758}.  Therefore, symptom-based detection of infection is less effective in COVID-19, compared to influenza and the SARS 2003 epidemic \cite{KKW2020}. To examine asymptomatic transmission is necessary to fully consider  in forecasting the spread size of the COVID-19 pandemic for effective public health prevention and control.

Driven by the nonignorable presence of asymptomatic transmission of the COVID-19 disease, we present a new dynamic model for the spread of infection by partitioning the infectious compartment of the traditional SIR model into three parts,  i.e., asymptomatic cases, pathologically infected but unconfirmed cases, and confirmed cases. 

Another crucial factor is dynamic human mobility in formulating the spread of the COVID-19 pandemic, driven by the effectiveness of isolating the spread of COVID-19 through lockdown of Wuhan city before the Chinese Spring Festival. The key initiative of such an extreme public health intervention  {ahead of} major public holidays was to cut off the massive human movement between Wuhan and surrounding cities. The spread of virus was controlled, with evidence that the reproductive number $R_t$ decreased over time $t$, by 2.7-3.8 before January 26 to less than 1.0 after February 6, and less than 0.3 after March 1 \cite{jama.2020.6130}. This encourages us to incorporate dynamic spreading patterns within the spatial framework of the human mobility network \cite{Brockmann1337}. 
The predictability of {the spread} of infectious disease could be improved by characterizing the geographical spread of epidemics  \cite{rvachev1985mathematical,hufnagel2004forecast}.
Population flowing out of Wuhan has been incorporated   to predict the risk and distribution of confirmed cases spatially \cite{Jia2020}. Geographical dispersion of an epidemic through human mobility has also been successfully applied to forecast the spreading of SARS 2003 and H1N1 2009 \cite{Brockmann1337}, whereas the mobility parameter was fixed as a constant on all nodes. Such constant mobility assumption is reasonable for SARS 2003 but not applicable for COVID-19 because the latter infectious disease is more contagious and there is much more intense human mobility in 2020 compared to 13 years ago \cite{Callaway2020nature}. 
 
In expanding the SIR model to our proposed SaucIR model, our main concerns focus on two aspects: (1) the division of infectious compartment into three separate segments, and (2) dynamic human mobility by separating the group size of migration in and out of a node within a network. It improves the prediction fidelity as shown in our simulation results. Furthermore, we demonstrate that it is possible to control the confirmed cases by reducing aggregation of migration among nodes in the human mobility network.

The remainder is organized as below. In Section 2, we give the formulation of  the new SaucIR model for the spread of COVID-19 based on its epidemiological characteristics and dynamic geographical spread. We present a practical way to measure dynamic human mobility  and suggest optimal enhancement for public control strategies. In Section 3, we apply the SaucIR model to {estimate and predict} the confirmed cases in mainland China and in worldwide countries with severe infections. In Section 4, we assess the predictive accuracy by comparing the proposed and the existing resembling models.

\section{Formulation of new spread model of COVID-19}

In this section, we {formulate} the predictive spread model of COVID-19 involving key factors driven from the epidemiological characteristics such as asymptomatic cases and dynamic human mobility {within a network of coupled populations. }We start from the following dynamics model \cite{Brockmann1337} describing the local disease time course on each node based on the conventional SIR model:
\begin{equation}\label{SIR}
\begin{split}
	& \partial_t S_n = -\frac{\alpha I_n S_n}{N_n}, \\
	& \partial_t I_n = \frac{\alpha I_n S_n}{N_n} - \beta_0 I_n,
\end{split}
\qquad n = 1,\dots,M,
\end{equation}
%{where $S_n$, $I_n$, and $R_n$ represent the numbers of susceptible, infected, and recovered individuals on a node $n$ in a network of people flow with totally $M$ nodes, respectively,}
where $S_n$ and $I_n$ represent the numbers of susceptible and infected individuals on node $n$ in a network of people flow with a total of $M$ nodes, respectively,
$N_n$ is the population size of node $n$,
$R_n=N_n-S_n-I_n$ represents the number of recovered or deceased on node $n$, $\alpha$ is the  mean infection rate of individuals, $\beta_0$ is the mean recovery rate of individuals, {and $\partial_t X_n$ stands for the partial derivative of the population size of state $X$ on node $n$ with respect to time $t$ in a time course with $X$ being the place holder for $S$, $I$, or any other disease state alphabet in the modeling thereafter.} The epidemiological threshold $R_0= \alpha/\beta$ is an indication of the transmissibility of a virus, governing the time evolution of the aforementioned equations.
% also known as the basic regeneration ratio

\subsection{Segmenting the infectious compartment}\label{sec2.1}

\begin{figure}[!h]
	\centering
	\includegraphics[width=0.5\textwidth,height=0.12\textheight]{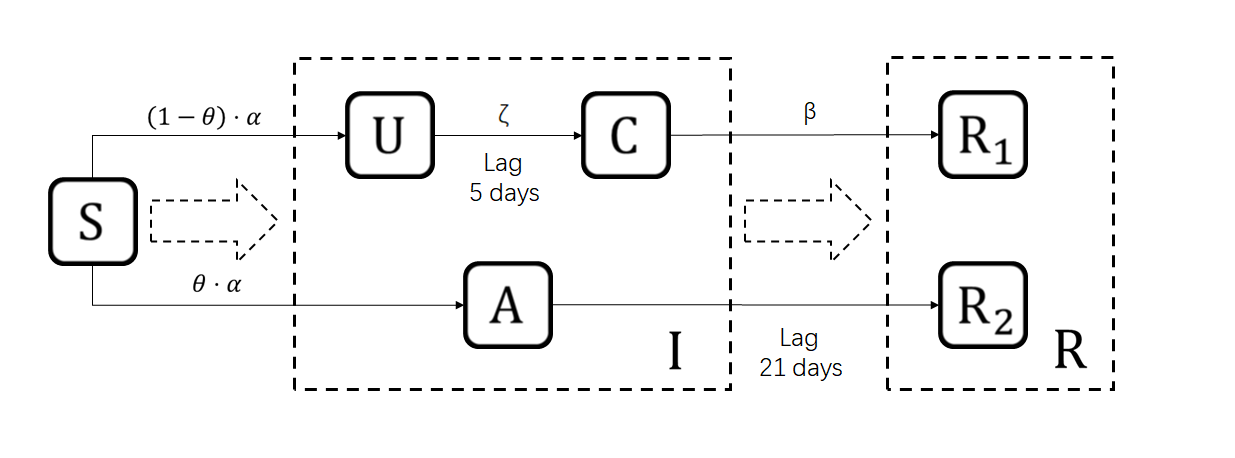}
	\caption{Progression flow of COVID-19.
	The parameters within the figure are illustrated in the adjacent paragraph.}
    \label{Progression}
\end{figure}

One of the epidemiological characteristics of COVID-19 that is remarkably different from existing infectious diseases such as SARS is the existence of asymptomatic carriers. Hence, symptom-based screening alone failed to detect a high proportion of infectious cases for asymptomatic persons and was not enough to control transmission. The probable asymptomatic transmission will not be known for sure unless an antibody test is applied. The risk of asymptomatic transmission cannot be ignored because of a large number of asymptomatic carriers \cite{mizumoto2020estimating,bai2020presumed,leejamainternmed2020}.
Thus, it is reasonable to segment the asymptomatic infected  COVID-19 carriers independently in the infectious compartment.  
As a result, the infectious compartment is partitioned into two segments, i.e., the asymptomatic ($A$) and the symptomatic. 
Additionally, the symptomatic segment exposes to dynamic process and is stratified into two stages, i.e., pathologically infected but unconfirmed  ($U$) and confirmed ($C$). That is, the infectious group first experiences the onset of symptoms, then it may be confirmed clinically after a time lag, called the incubation period. Thereafter, the confirmed group  %a lag of average five days of incubation \citep{lauer2020incubation},  
will be isolated for therapy in the hospital and lose its transmissibility. The incubation period is contagious in our compartmental model \cite{Li2020IJID}, which is in line with the existing assumption for modeling the spread of the SARS epidemic \cite{chowell2003sars}.
The infectivity during the incubation period for COVID-19 is critical for controlling the disease \cite{Liu2020jtm}.
As a result of the newly segmented infectious compartment, the $R$ (recovered or deceased) compartment composes of two parts, recovered or deceased ($R_1$) from the symptomatic compartment or recovered ($R_2$) from the asymptomatic compartment accordingly. 
The formulated contagion pattern is summarized in  
Figure \ref{Progression} to illustrate the spread dynamics.

Next, we analyze the key epidemiological parameters in Figure \ref{Progression}.  The parameter $\alpha$ denotes the transmission rate from susceptible to infected, $\theta$ denotes the probability that susceptible individuals turn to be asymptomatically infected, $\zeta$ denotes the rate that a person in the incubation period turns to be confirmed clinically, and $\beta$ denotes the recovery rate from symptomatically confirmed to recovered or deceased. 
For the symptomatic segment, the time lag from $U$ to $C$ is taken as 5 days, the average incubation period in the literature \cite{lauer2020incubation}. Our simulations show that the incubation period of 3 to 5 days is not sensitive to the proposed model.   
For the progression from asymptomatic ($A$) to recovered ($R_2$), the time lag 21 days corresponds to the maximal observation of the communicable period of asymptomatic  carriers, 
counted from the first day of positive nucleic acid test to the first day of the continuous negative tests of an asymptomatic carrier \cite{bai2020presumed}. There are very few references for the interval length of the communicable period of asymptomatic carriers to the pandemic spread \cite{hu2020clinical}. Our simulation results show that the scope of 9-21 days will not affect much of the accuracy of the proposed model in Section 3.1.
The aforementioned consideration of epidemiological characteristics is summarized in the following compartmental model on node $n$,
\begin{equation}\label{eqnInitial}
\begin{split}
& \partial_t S_n = -\frac{\alpha_n (U_n+A_n)S_n}{N_n}, \\
& \partial_t U_n = \frac{\alpha_n (U_n+A_n)S_n}{N_n}(1-\theta)-\zeta_n U_n^{t-5}, \\
& \partial_t A_n = \frac{\alpha_n (U_n+A_n)S_n}{N_n}\theta-A_n^{t-21}, \\
& \partial_t C_n = \zeta_n U_n^{t-5}-\beta_n C_n, \\
& \partial_t R_{1n} = \beta_n C_n, \\
& \partial_t R_{2n} = A_n^{t-21}, \qquad n = 1,\dots,M,
\end{split}
\end{equation}
where $A_n$ represents the group size of asymptomatic cases,  $U_n$ and $C_n$ are group sizes of infected but not yet confirmed clinically and confirmed clinically   in the symptomatic segment, respectively,  $R_{1n}$ and $R_{2n}$ denote the removed cases from symptomatic and asymptomatic segments, respectively, $\alpha_n$ denotes the average infection rate for population  $n$ in the network, $\beta_n$  denotes the average recovery rate of individuals from confirmed clinically under therapy  to recovered on node $n$, and $R_{1n} =N_n-S_n-(U_n+A_n+C_n)-R_{2n}$. 

Notice that for population $n$ in a network, the accumulated confirmed cases, denoted as $D$, is the sum of the clinically confirmed under therapy  ($C$) and the recovered or deceased from symptomatic cases  ($R_1$). Thereafter,
\begin{equation}\label{D=C+R1}
\partial_t D_n = \zeta_n U_n^{t-5}
\end{equation}
will replace the bi-interrelated variation equations in the system  of equations (\ref{eqnInitial}) in the final dynamic model.

\subsection{Dynamic human mobility}
We now incorporate the impact of population ﬂow onto the modeling of the infectious spread of COVID-19. To guarantee the prediction ﬁdelity, we distinguish the group size of people emanating from a node $m$ in and out of another node $n$, and hence, the distinct human mobility rates of moving in and out of a node $n$. This is a quite practical strategy considering the diﬀerence among nodes. Let $X$ be a placeholder for the compartments $S$, $U$, and $A$. The infection variation for population $n$ comes from two sources, local and migration,
\begin{equation}\label{plusMobility}
\partial_t X_n=\partial_t X_{n,\textrm{local}}+\partial_t X_{n,\textrm{migration}},
\end{equation}
where the first summand   $\partial_t X_{n,\textrm{local}}$ on the right side of equation (\ref{plusMobility}) represents the dynamic local disease time course on  node $n$ by equations (\ref{eqnInitial}) and (\ref{D=C+R1}). Next, we will elucidate the formulating of $\partial_t X_{n,\textrm{migration}}$ based on parameters of dynamic human mobility.

Let $F_{nm}$ be the group size of people emanating from node $m$ to node $n$. Denote $F_{n \cdot} = \sum_m F_{nm}$ and $F_{\cdot m} =\sum_n F_{nm}$. Let $\gamma^{\textrm{in}}_{n}=F_{n \cdot}/N_n$ be the proportion of the group size moving in destination  $n$ {($F_{n \cdot}$)} to the population size of node $n$ ($N_n$), and $\gamma^{\textrm{out}}_{n}=F_{\cdot n}/N_n$ be the proportion of group size of people leaving  node $n$ ($F_{\cdot n}$) to the population size of node $n$. Let {$P^{\textrm{\textrm{in}}}_{nm}= F_{nm}/F_{n \cdot}$} be the proportion of group size emanating from node $m$ in destination  $n$ 
%{($F_{nm}$)} 
to the group size moving in destination $n$; %{($F_{n \cdot}$)}, 
let $P^{\textrm{out}}_{mn}=F_{mn}/F_{\cdot n}$ be the proportion of group size leaving node $n$ for another node  $m$ %{($F_{mn}$)} 
to the group size emanating from  node $n$. %{($F_{\cdot n}$)}. 
Let $P_{nn}^{\textrm{\textrm{in}}}=0=P_{nn}^{\textrm{out}}$.
Under the setting of  such human mobility measurements, the variation of the susceptible within our spread model of COVID-19 satisfies
\begin{equation}\label{equationS}
\begin{split}
&    \partial_t S_{n,\textrm{migration}} = \\
&    \sum_{m=1}^{M}\bigg[\gamma_{n}^{\textrm{\textrm{in}}}P_{nm}^{\textrm{\textrm{in}}}S_m - \gamma_{n}^{\textrm{\textrm{in}}}P_{nm}^{\textrm{\textrm{in}}}\frac{\alpha_m(U_m+A_m)S_m}{N_m}\\
&\quad \quad    -\gamma_{n}^{\textrm{out}}P_{mn}^{\textrm{out}}S_n \bigg].\\
    \end{split}
\end{equation}
Notice that, when there is no geometrical difference in transmission, denoted as $\gamma^{\textrm{\textrm{in}}}_{n}=\gamma^{\textrm{out}}_{n} \equiv \gamma$ for any node $n$, and there is no infection during the migration, denoted as $\alpha_m=0$, our proposed 
equation (\ref{equationS})  reduces to the representation of $\partial_t S_{n}$ in the literature \cite{Brockmann1337}. 

For $U_n$ and $A_n$, one cannot ignore the infection owing to human mobility among nodes in the geometrical network. The correspondent dynamic processes can be presented as follows:
\begin{equation}\label{equationU&A}
\begin{split}
	& \partial_t U_{n,\textrm{migration}}\\
= &\sum_{m=1}^M\bigg[\gamma_{n}^{\textrm{\textrm{in}}}P_{nm}^{\textrm{\textrm{in}}}U_m +\gamma_n^{\textrm{\textrm{in}}}P_{nm}^{\textrm{\textrm{in}}}\frac{\alpha_m(1-\theta)(U_m+A_m)S_m}{N_m}  \\
&\quad \quad -\gamma_{n}^{\textrm{out}}P_{mn}^{\textrm{out}}U_n\bigg],\\
& \partial_t A_{n,\textrm{migration}} \\
=& \sum_{m=1}^{M}\bigg[\gamma_{n}^{\textrm{\textrm{in}}}P_{nm}^{\textrm{\textrm{in}}}A_m +\gamma_n^{\textrm{\textrm{in}}}P_{nm}^{\textrm{\textrm{in}}}\frac{\alpha_m\theta(U_m+A_m)S_m}{N_m}\\
&\quad \quad - \gamma_{n}^{\textrm{out}}P_{mn}^{\textrm{out}}A_n\bigg],\end{split}
\end{equation}
where $\alpha_m$ denotes the rate of infection at {node} $m$ other than node $n$.

Based on the spirit of equation (\ref{plusMobility}), we can further update the dynamic model (\ref{eqnInitial}) by combining (\ref{D=C+R1}), (\ref{equationS}), and (\ref{equationU&A}) into (\ref{eqnInitial}). % we have
%\begin{equation}\label{equation5}
%\begin{split}
% \partial_t S_n =& -\frac{\alpha_n (U_n+A_n)S_n}{N_n} \\
%& + \sum_{m \neq n}(\gamma_n^{\textrm{\textrm{in}}}P_{nm}^{\textrm{\textrm{in}}}S_m - \gamma_{n}^{\textrm{\textrm{in}}}P_{nm}^{\textrm{\textrm{in}}}\frac{\alpha_m(U_m+A_m)S_m}{N_m}\\&\qquad\quad- \gamma_n^{\textrm{out}}P_{mn}^{\textrm{out}}S_n),  \\
% \partial_t U_n = &\frac{\alpha_n (U_n+A_n)S_n}{N_n}(1-\theta)-\zeta_n U_n^{t-5} \\
%& + \sum_{m \neq n}(\gamma_n^{\textrm{\textrm{in}}}P_{nm}^{\textrm{\textrm{in}}}U_m+ \gamma_n^{\textrm{\textrm{in}}}P_{nm}^{\textrm{\textrm{in}}}\frac{\alpha_m(1-\theta)(U_m+A_m)S_m}{N_m} \\
%&\qquad\quad - \gamma_n^{\textrm{out}}P_{mn}^{\textrm{out}}(U_n)),\\
% \partial_t A_n = &\frac{\alpha_n (U_n+A_n)S_n}{N_n}\theta-A_n^{t-21} \\
%& + \sum_{m \neq n}(\gamma_n^{\textrm{\textrm{in}}}P_{nm}^{\textrm{\textrm{in}}}A_m+\gamma_n^{\textrm{\textrm{in}}}P_{nm}^{\textrm{\textrm{in}}}\frac{\alpha_m\theta(U_m+A_m)S_m}{N_m} \\
%&\qquad\quad - \gamma_n^{\textrm{out}}P_{mn}^{\textrm{out}}(A_n)),\\
% \partial_t D_n =&\zeta_n U_n^{t-5}, \\
% \partial_t R_{2n} =& A_n^{t-21}, \qquad n = 1,\dots,M,
%\end{split}
%\end{equation}

\subsection{The SaucIR-model}
To formulate the final SaucIR model, we still need to further refine some nested parameters to enhance predictivity. In a network, the prevention and control policies of different nodes should lead to spatially varying epidemic parameters.

First of all, we will add the {effects} of quarantine  into the modeling of {the} contagion pattern of COVID-19 \cite{chowell2003sars}. 
In mainland China, during the COVID-19 pandemic, most provinces have announced officially the detailed confirmed cases and labeled whether the specific case was once quarantined. This information is beneficial for estimating the quarantine rate parameter $l_n$  as the ratio of the size of quarantine labeled cases against the total number of confirmed cases of COVID-19 disease for population $n$. 
Still in modeling the spread of SARS by 
the literature \cite{chowell2003sars}, $l_n$ changes over time. 
Varying $l_n$ is effective in the situation where there exist inflection points, validated by Figure 4 in literature \cite{chowell2003sars}, whereas it is not significant for our proposed model, as confirmed by the simulations summarized in Table \ref{tlmape} and Figure \ref{provincePrediction}.

The other focus of our study is the correction of the transmission rate $\alpha_m$ for the migration process.
The transmission rate function is no doubt time and nodal dependent. Let $\tau_n$ be a decay constant on node $n$. The general exponential tilt expression
\begin{equation}\label{exptilt}
\alpha_n(t)= \alpha_n(0) \cdot e^{-\tau_n t}
\end{equation}
is an acceptable functional to act in the local infectious process \cite{chen2020data} in which the prevention and control in local governments can effectively help prevent the spread of contagion. However, the monotone decreasing function (\ref{exptilt}) is not suitable to apply in the migration process owing to unprecedented human mobility. Conservatively, we uniformly apply the nodal function $\alpha_n(t)$ to behave as the transmission rate function in the migration part of the final model.

All of the above considerations lead to the final SaucIR model illustrating the infection  progression:
\begin{equation}\label{SaucIR}
\begin{split}
& \partial_t S_n = -(1-l_n)\frac{\alpha_n(t) (U_n+A_n)S_n}{N_n} \\
& \qquad \quad + \sum_{m=1}^M[\gamma_n^{\textrm{\textrm{in}}}P_{nm}^{\textrm{\textrm{in}}}S_m\\
& \qquad \quad- \gamma_{n}^{\textrm{\textrm{in}}}P_{nm}^{\textrm{\textrm{in}}}(1-l_n)\frac{\alpha_m(0)(U_m+A_m)S_m}{N_m}\\&\qquad\quad - \gamma_n^{\textrm{out}}P_{mn}^{\textrm{out}}S_n],  \\
& \partial_t U_n = (1-l_n)\frac{\alpha_n(t) (U_n+A_n)S_n}{N_n}(1-\theta)-\zeta_n U_n^{t-5} \\
& \qquad \quad  + \sum_{m=1}^M[\gamma_n^{\textrm{\textrm{in}}}P_{nm}^{\textrm{\textrm{in}}}(1-l_n)U_m\\
& \qquad \quad+\gamma_n^{\textrm{\textrm{in}}}P_{nm}^{\textrm{\textrm{in}}}(1-l_n)\frac{\alpha_m(0)(1-\theta)(U_m+A_m)S_m}{N_m} \\
& \qquad \qquad\qquad - \gamma_n^{\textrm{out}}P_{mn}^{\textrm{out}}(1-l_n)U_n], \\
& \partial_t A_n =(1-l_n)\frac{\alpha_n(t) (U_n+A_n)S_n}{N_n}\theta-A_n^{t-21} \\
& \qquad \quad+ \sum_{m=1}^M[\gamma_n^{\textrm{\textrm{in}}}P_{nm}^{\textrm{\textrm{in}}}(1-l_n)A_m\\
& \qquad \quad+\gamma_n^{\textrm{\textrm{in}}}P_{nm}^{\textrm{\textrm{in}}}(1-l_n)\frac{\alpha_m(0)\theta(U_m+A_m)S_m}{N_m}\\
& \qquad \qquad\qquad-\gamma_n^{\textrm{out}}P_{mn}^{\textrm{out}}(1-l_n)A_n],\\
& \partial_t D_n = \zeta_n U_n^{t-5}, \\
& \partial_t R_{2n} = A_n^{t-21}, \qquad n = 1,\dots,M.
\end{split}
\end{equation}
The epidemic parameters $l_n$, $\alpha_n$, and $\zeta_n$ are estimated from the announced epidemic data by the provinces.  The initial value of $S_n$ is the population size of every node. The initial value of $D_n$ is the number of cumulative confirmed cases each day. The initial value of $U_n$ is the sum of new diagnoses in the next six days. The initial value of $A_n$ is the total number of new diagnoses in the next six days multiplied by $\theta/(1-\theta)$.

\subsection{Prevention and control strategy}

In this subsection, we explore whether human mobility can be modified among nodes in a network to control or minimize the accumulated confirmed cases in a district over a certain time interval. 
 
Let $D_{n}^T$ be the confirmed cases on node $n$ up to day $T$. Denote any local coupled populations in the network as $S=\{1,\cdots, K\}$, and all populations in the network as $S_0=\{1,\cdots, M\}$, where $K \leq M$.
The migration-related parameters introduced before Section 2.3 on a specific day $t$ are then denoted as  $\gamma_{n,t}^{\textrm{\textrm{in}}}$, $\gamma_{m,t}^{\textrm{out}}$, $P_{nm,t}^{\textrm{\textrm{in}}}$, and $P_{nm,t}^{\textrm{out}}$.
The objective function will be optimized by choosing the appropriate time-dependent tuning parameters aforementioned.
 
Recall that we define $\gamma^{\textrm{in}}_n$, $\gamma^{\textrm{out}}_n$, $P^{\textrm{\textrm{in}}}_{nm}$, and $P^{\textrm{\textrm{out}}}_{mn}$, the parameters of the human mobility in a network, in Section 2.2. We add an extra suffix $t$ in the footnote to denote the corresponding parameters when time $t$ falls into the time interval $[0, T]$.
Let $C_{nm, T}^{\textrm{\textrm{in}}}$ and $C_{nm, T}^{\textrm{out}}$ be the rate of emanating  from  node $m$ in and out to node $n$, respectively, in the network over the time period $[0, T]$. Then, it is readily seen that
\begin{equation}\label{Cmn_in_out}
\begin{split}
%&\min_{\gamma_{n}^{\textrm{\textrm{in}}},P_{nm}^{\textrm{\textrm{in}}},\gamma_{n}^{\textrm{out}},P_{mn}^{\textrm{out}}}\sum_n D_{n}^T \text{ subject to }\\
&\sum_{t\in [0,T]}\gamma_{n,t}^{\textrm{\textrm{in}}} P_{nm,t}^{\textrm{\textrm{in}}}=C_{nm,T}^{\textrm{\textrm{in}}},\\
&\sum_{t\in [0,T]}\gamma_{m,t}^{\textrm{out}} P_{nm,t}^{\textrm{out}}=C_{nm,T}^{\textrm{out}},\ \forall m,n \in S_0.\ \
%& {C_{nm}^{\textrm{\textrm{in}}}N_n=C_{nm}^{\textrm{out}}N_m, \ \forall m,n.}
\end{split}
\end{equation}

To modify parameters of human mobility among nodes in the network, for any pairwise nodes $n$ and $m$, one natural constraint is to keep a constant relationship between the group size emanating from node $n$ with population size $N_n$ and out of another node $m$ with population size $N_m$, so that the optimization problem may be presented as
\begin{equation}\label{optimal}
\begin{split}
&\min_{\gamma_{n,t}^{\textrm{\textrm{in}}},P_{nm,t}^{\textrm{\textrm{in}}},\gamma_{m,t}^{\textrm{out}},P_{nm,t}^{\textrm{out}}}\sum_{n \in S}D_{n}^T \text{ subject to }\\
%&\sum_{t\in [0,T]}\gamma_{n}^{\textrm{\textrm{in}}} P_{nm}^{\textrm{in}}=C_{nm}^{\textrm{in}}\text{ and } \sum_{t\in [0,T]}\gamma_{n}^{\textrm{out}} P_{mn}^{\textrm{out}}=C_{mn}^{\textrm{out}},\ \forall m,n\\
& {C_{nm,T}^{\textrm{in}}N_n=C_{nm,T}^{\textrm{out}}N_m, \ \forall m,n \in S_0.}
\end{split}
\end{equation}
For the optimization problem (\ref{optimal}), we assume that the mobility group will not change the epidemic parameters $\alpha_n(t)$, $\beta_n$, $l_n$, and $\theta$ on node $n$. To solve the optimization problem (\ref{optimal}), we employ the general genetic algorithm \cite{mitchell1998introduction}. Here are notations involved in the algorithm. 
Denote $(\gamma P)^{\textrm{in}}_{nm,t}=\gamma_{n,t}^{\textrm{in}} P_{nm,t}^{\textrm{in}}$ and $(\gamma P)_{mn,t}^{\textrm{out}}=\gamma_{m,t}^{\textrm{out}} P_{nm,t}^{\textrm{out}}$; 
let $C_T^{\textrm{in}}$ and $C_T^{\textrm{out}}$ represent the matrices $(C_{nm,T}^{\textrm{in}})_{n, m \in S_0}$ and $(C_{nm,T}^{\textrm{out}})_{n, m \in S_0}$, respectively; and let $(\gamma P)^{\textrm{in}}$ and $(\gamma P)^{\textrm{out}}$ represent the three-dimensional matrices $((\gamma P)^{\textrm{in}}_{nm,t})_{t\in[0, T],n, m \in S_0}$   and $((\gamma P)_{mn,t}^{\textrm{out}})_{t\in[0, T],n, m \in S_0}$, respectively. 
The particular terminologies in the genetic algorithm and their correspondence in our problem are listed as below:

\begin{description}

\item 1.  Individual: a matrix shaped as (T, M, M), where T and M correspond to $(\gamma P)^{\textrm{in}}_{nm,t}$ and $(\gamma P)_{mn,t}^{\textrm{out}}$, respectively, for $t\in [0, T]$ and $n, m \in S_0$.

\item 2. Population: a collection of individuals. 

%3. Loop time, that how many times we do the loop.
\item 3. Fitness: a description of how good an individual fits the environment, and in our study, how small the number of $\sum_n D_n^T$ is. 

\item 4. Selection, crossover, and mutation: specific operation in genetic algorithm, aiming to generate individuals that are more likely to correspond to better fitness.
\end{description} 

Algorithm 1 provides the steps for minimizing the accumulated confirmed cases within any sub-network over a time interval, of which the function \textsl{getfitness}  is presented in the following Algorithm 2.
\begin{table}[h]\label{algorithm1}
\begin{tabular}{ l }
\hline
\leftline{Algorithm 1: minimize the accumulated confirmed cases}\\
\hline
%\leftline{\textbf{Require}: population size, C^{\textrm{in}}, C^{\textrm{out}}, loop num, T, N\\}
\leftline{1: Initialize population}\\
\leftline{2: for iter = 1,$\ldots$, loop num do}\\
\leftline{3: \quad for individual in population do}\\
\leftline{4: \quad \quad fitness(individual)=getfitness($C_T^{\textrm{in}}$,$C_T^{\textrm{out}}$,individual,$T$,$M$)}\\
\leftline{5: \quad end for}\\
%\leftline{6:     current-parameters, current-fitness = best(population,fitness)\\}
%\leftline{7:     if current-fitness \leq best-fitness\\}
%\leftline{8:            best-fitness = current-fitness\\}
%\leftline{9:            best-parameters = current-parameters\\}
\leftline{6: \quad record the best fitness so far}\\
%\leftline{10:    end if}
\leftline{7: \quad selection(population, fitness)}\\
\leftline{8: \quad  crossover(population)}\\
\leftline{9: \quad mutation(population)}\\
\leftline{10: end for}\\
\hline
\end{tabular}
\end{table}

%are in short for $\gamma_{n,t}^{\textrm{in}} P_{nm,t}^{\textrm{in}}$ and $\gamma_{m,t}^{\textrm{out}} P_{nm,t}^{\textrm{out}}$, respectively;
%And in algorithm 1, we summarize the process of using the SaucIR model to get fitness , a description of how good an individual fit the environment, in our problem, how small the number of $\sum_n D_n^T$ is. We use "model8" to express the process of get  $\sum_n D_n^T$ with a set of  $\gamma P^{\textrm{in}}_{nm}$   and $\gamma P_{mn}^{\textrm{out}}$. In algorithm 2, we show how to minimize the accumulated confirmed cases. 
%Here $t\in[0, T]$ and $n, m \in S_0$. The algorithm is listed in Algorithm 1.}
 
 \begin{table}[h]\label{algorithm2}
 \begin{tabular}{ l }
\hline
\leftline{Algorithm 2: function \textsl{getfitness} implementing the SaucIR model}\\
\hline
\leftline{\textbf{input}: $C^{\textrm{in}}$, $C^{\textrm{out}}$, individual, $T$, $M$}\\
\leftline{1: set individual($t$, $m$, $m$)=0}\\
\leftline{2: for $m$ in range($M$) do}\\
\leftline{3: \quad for $n$ in range($M$) do}\\
\leftline{4: \quad \quad sum$_{mn}$=$\sum_t$(individual($t,m, n$))}\\
\leftline{5: \quad \quad for $t$ in range($T$) do}\\
\leftline{6:  \quad \quad \quad $(\gamma P)^{\textrm{in}}_{mn,t}$ = individual($t, m, n)C_{mn,T}^{\textrm{in}}$/sum$_{mn}$}\\
\leftline{7: \quad \quad \quad $(\gamma P)^{\textrm{out}}_{mn,t}$ =
 individual($t, m, n$)$C_{mn,T}^{\textrm{out}}$/sum$_{mn}$}\\
\leftline{8: \quad \quad end for}\\
\leftline{9: \quad end for}\\
\leftline{10: end for}\\
\leftline{11: fitness=$-$equations (8)$((\gamma P)^{\textrm{in}},(\gamma P)^{\textrm{out}})$}\\
\hline
\end{tabular}
\end{table}
Based on the SaucIR model (\ref{SaucIR}), one may use Algorithms 1 and 2 to solve the minimization problem (\ref{optimal}). Notice that the four tuning rate parameters are ratios among $F_{\cdot m}$, $F_{n \cdot}$, $N_n$,  $N_m$, and $F_{nm}$. Therefore, governmental management may have a proper intervention on the migration size within a sub-network for the purpose of preventing and controlling the spread of COVID-19.

%To obtain the optimal solution, we suppose that the mobility group will not change the epidemic parameters $\alpha_n(t)$, $\beta_n$, $l_n$ and $\theta$ on node $n$ up to time $t$. Practically, $\gamma_{n}^{\textrm{in}} P_{nm}^{\textrm{in}}$ and $\gamma_{n}^{\textrm{out}} P_{mn}^{\textrm{out}}$ are treated as an integrated single parameter for computing convenience. The general genetic algorithm is used to search the optimal solutions for the minimizing problem (\ref{optimal}) with the spirit of conducting cross validation on the adaptation function with a large magnitude and disposing of the low-level adaptation function \cite{mitchell1998introduction}.

\section{Numerical analysis and forecasting}
In this section, we use the data of confirmed cases within the human mobility networks in mainland China and the international community to analyze the predictive fidelity of the SaucIR spread model (\ref{SaucIR}) of COVID-19. To assess the accuracy of prediction, we  adopt two assessment indices, max absolute percent error (MAPE) and root mean squared error (RMSE) with mathematical expressions 
$$
\text{MAPE}_n=\max_t\bigg\{\frac{|D_{n}^{\text{pred}}-D_{n}^{\text{real}}|}{D_{n}^{real}}\bigg\}
$$
and 
$$\text{RMSE}_n=\sqrt{\sum_t(D_{n}^{\text{pred}}-D_{n}^{\text{real}})^2/(t_0-1)},
$$
respectively, where $t_0$ is the number of days of a prediction period with daily unity to count time, and the superscripts ``$\text{pred}$'' and ``$\text{real}$'' denote predicted and real, respectively. MAPE reflects the maximal daily percentage error out of the total number of infections over the forecasting time period, whereas RMSE measures the averaged headcount error over the forecasting time period. The prediction is more accurate with lower levels of such errors.
%max error and error as a proportion of total number of infections)

\subsection{Analysis of human mobility network in mainland China} \label{sec3.1}

\begin{table*}[h!]
\caption{MAPE assessing effect of asymptomatic infections spatially }
\label{t1}
\begin{center}
\begin{tabular}{ llllllllllll }
  \hline
  &\multicolumn{11}{c}{Province} \\
  \cline{2-12}
  $\theta$ &\footnotesize Beijing &\footnotesize Shanghai &\footnotesize Jiangsu &\footnotesize Zhejiang &\footnotesize Anhui &\footnotesize Guangdong &\footnotesize Henan &\footnotesize Hunan &\footnotesize Chongqing &\footnotesize Sichuan &\footnotesize Jiangxi\\
  \hline
  0.05 &0.033&0.033&0.028&0.019&0.008&0.032&0.008&0.018&0.012&0.023&0.025\\
  0.15 &0.013&0.021&0.019&0.017&0.006&0.009&0.004&0.022&0.009&0.008&0.004\\
  0.25 &0.017&0.006&0.015&0.011&0.006&0.015&0.005&0.014&0.009&0.009&0.010\\
  0.35 &0.015&0.006&0.014&0.006&0.005&0.015&0.003&0.003&0.005&0.007&0.017\\
  0.45 &0.015&0.006&0.019&0.009&0.001&0.028&0.011&0.012&0.003&0.005&0.012\\
  0.55 &0.026&0.021&0.011&0.029&0.007&0.047&0.008&0.009&0.012&0.023&0.028\\
  \hline
\end{tabular}
\end{center}
\end{table*}

\begin{table*}[h!]
\caption{RMSE assessing effect of asymptomatic infections spatially}
\label{t2}
\begin{center}
\begin{tabular}{ llllllllllll }
  \hline
  &\multicolumn{11}{c}{Province} \\
  \cline{2-12}
  $\theta$ &\footnotesize Beijing &\footnotesize Shanghai &\footnotesize Jiangsu &\footnotesize Zhejiang &\footnotesize Anhui &\footnotesize Guangdong &\footnotesize Henan &\footnotesize Hunan &\footnotesize Chongqing &\footnotesize Sichuan &\footnotesize Jiangxi\\
  \hline
  0.05 &11&13&15&22&6&48&8&20&8&13&23\\
  0.15 &5&7&10&19&6&15&6&23&5&4&3\\
  0.25 &5&2&8&10&5&17&7&13&5&5&8\\
  0.35 &5&2&7&8&5&17&3&3&3&4&13\\
  0.45 &5&2&10&9&1&30&16&11&2&2&10\\
  0.55 &9&9&16&31&6&56&10&8&6&12&24\\
  \hline
\end{tabular}
\end{center}
\end{table*}

\begin{table*}[h!]
\caption{{MAPE assessing effect of different lag of days of incubation period}}
\label{t5daysmape}
\begin{center}
\begin{tabular}{ llllllllllll }
  \hline
  &\multicolumn{11}{c}{Province} \\
  \cline{2-12}
  \footnotesize Lag (days) &\scriptsize Beijing &\scriptsize Shanghai &\scriptsize Jiangsu &\scriptsize Zhejiang &\scriptsize Anhui &\scriptsize Guangdong &\scriptsize Henan &\scriptsize Hunan &\scriptsize Chongqing &\scriptsize Sichuan &\scriptsize Jiangxi\\
  \hline
  3 &0.018&0.006&0.019&0.008&0.008&0.017&0.012&0.021&0.018&0.010&0.026\\
  4 &0.016&0.006&0.017&0.007&0.007&0.008&0.010&0.002&0.002&0.004&0.010\\
  5 &0.018&0.006&0.016&0.011&0.006&0.015&0.006&0.015&0.009&0.010&0.011\\
  6 &0.015&0.012&0.029&0.004&0.012&0.011&0.006&0.007&0.007&0.019&0.012\\
  7 &0.043&0.024&0.021&0.016&0.021&0.012&0.029&0.016&0.016&0.041&0.016\\
  \hline
\end{tabular}
\end{center}
\end{table*}

\begin{table*}[h!]
\caption{{Range of uncertainty of MAPE assessing effect of [3,5] lag of days of incubation period}}
\label{t5daysmaperange}
\begin{center}
\begin{tabular}{ llllllllllll }
  \hline
  &\multicolumn{11}{c}{Province} \\
  \cline{2-12}
   &\scriptsize Beijing &\scriptsize Shanghai &\scriptsize Jiangsu &\scriptsize Zhejiang &\scriptsize Anhui &\scriptsize Guangdong &\scriptsize Henan &\scriptsize Hunan &\scriptsize Chongqing &\scriptsize Sichuan &\scriptsize Jiangxi\\
  \hline 
\footnotesize Maximum &0.018&0.006&0.019&0.011&0.008&0.017&0.012&0.021&0.018&0.01&0.026\\
\footnotesize Mininum &0.016&0.006&0.016&0.007&0.006&0.008&0.006&0.002&0.002&0.004&0.01\\
  \hline
\end{tabular}
\end{center}
\end{table*}

\begin{table*}[h!]
\caption{{RMSE assessing effect of different lag of days of incubation period}}
\label{t5daysrmse}
\begin{center}
\begin{tabular}{ llllllllllll }
  \hline
  &\multicolumn{11}{c}{Province} \\
  \cline{2-12}
  \footnotesize Lag (days) &\scriptsize Beijing &\scriptsize Shanghai &\scriptsize Jiangsu &\scriptsize Zhejiang &\scriptsize Anhui &\scriptsize Guangdong &\scriptsize Henan &\scriptsize Hunan &\scriptsize Chongqing &\scriptsize Sichuan &\scriptsize Jiangxi\\
  \hline
  3 &5&1&9&8&6&21&13&19&9&4&20\\
  4 &4&2&8&7&7&9&11&2&1&2&7\\
  5 &5&2&8&10&4&17&6&12&4&4&8\\
  6 &5&4&15&5&8&10&6&7&4&7&12\\
  7 &14&9&12&22&19&12&37&18&7&22&17\\
  \hline
\end{tabular}
\end{center}
\end{table*}

\begin{table*}[h!]
\caption{Range of uncertainty of RMSE assessing effect of [3,5] lag of days of incubation period}
\label{t5daysrmserange}
\begin{center}
\begin{tabular}{ llllllllllll }
  \hline
  &\multicolumn{11}{c}{Province} \\
  \cline{2-12}
   &\scriptsize Beijing &\scriptsize Shanghai &\scriptsize Jiangsu &\scriptsize Zhejiang &\scriptsize Anhui &\scriptsize Guangdong &\scriptsize Henan &\scriptsize Hunan &\scriptsize Chongqing &\scriptsize Sichuan &\scriptsize Jiangxi\\
  \hline  
\footnotesize Maximum &5&2&9&10&7&21&13&19&9&4&20\\
\footnotesize Minimum &4&1&8&7&4&9&6&2&1&2&7\\
  \hline
\end{tabular}
\end{center}
\end{table*}

\begin{table*}[h!]
\caption{MAPE assessing effect of different lag of days of the median communicable period}
\label{t21daysmape}
\begin{center}
\begin{tabular}{ llllllllllll }
  \hline
  &\multicolumn{11}{c}{Province} \\
  \cline{2-12}
\footnotesize Lag (days) &\scriptsize Beijing &\scriptsize Shanghai &\scriptsize Jiangsu &\scriptsize Zhejiang &\scriptsize Anhui &\scriptsize Guangdong &\scriptsize Henan &\scriptsize Hunan &\scriptsize Chongqing &\scriptsize Sichuan &\scriptsize Jiangxi\\
  \hline
  21 &0.018&0.006&0.016&0.011&0.006&0.015&0.006&0.015&0.009&0.01&0.011\\
  18 &0.015&0.015&0.021&0.006&0.007&0.004&0.006&0.008&0.011&0.019&0.008\\
  15 &0.02&0.018&0.013&0.003&0.012&0.004&0.013&0.001&0.013&0.014&0.003\\
  12 &0.018&0.021&0.014&0.013&0.007&0.009&0.004&0.008&0.016&0.014&0.003\\
  9 &0.018&0.012&0.016&0.003&0.012&0.01&0.009&0.016&0.011&0.012&0.013\\
  6 &0.02&0.006&0.035&0.017&0.016&0.025&0.034&0.015&0.02&0.016&0.021\\
  \hline
\end{tabular}
\end{center}
\end{table*}

\begin{table*}[h!]
\caption{Range of MAPE assessing effect of [9,21] lag of days of the median communicable period}
\label{t21daysmaperange}
\begin{center}
\begin{tabular}{ llllllllllll }
  \hline
  &\multicolumn{11}{c}{Province} \\
  \cline{2-12}
   &\scriptsize Beijing &\scriptsize Shanghai &\scriptsize Jiangsu &\scriptsize Zhejiang &\scriptsize Anhui &\scriptsize Guangdong &\scriptsize Henan &\scriptsize Hunan &\scriptsize Chongqing &\scriptsize Sichuan &\scriptsize Jiangxi\\
  \hline
\footnotesize Maximum&0.02&0.021&0.021&0.013&0.012&0.015&0.013&0.016&0.016&0.019&0.013\\
\footnotesize Minimum&0.015&0.006&0.013&0.003&0.006&0.004&0.004&0.001&0.009&0.01&0.003\\
  \hline
\end{tabular}
\end{center}
\end{table*}

\begin{table*}[h!]
\caption{{RMSE assessing effect of different lag of days of the median communicable period}}
\label{tday21rmse}
\begin{center}
\begin{tabular}{ llllllllllll }
  \hline
  &\multicolumn{11}{c}{Province} \\
  \cline{2-12}
\footnotesize Lag (days) &\scriptsize Beijing &\scriptsize Shanghai &\scriptsize Jiangsu &\scriptsize Zhejiang &\scriptsize Anhui &\scriptsize Guangdong &\scriptsize Henan &\scriptsize Hunan &\scriptsize Chongqing &\scriptsize Sichuan &\scriptsize Jiangxi\\
  \hline
  21 &5&2&8&10&4&17&6&12&4&4&8\\
  18 &5&5&11&8&6&5&6&9&5&9&7\\
  15 &6&6&8&3&10&4&15&1&6&6&2\\
  12 &6&7&10&17&6&10&4&8&8&6&2\\
  9 &5&3&8&3&10&11&12&18&5&5&11\\
  6 &6&2&17&23&16&26&38&16&9&6&18\\
  \hline
\end{tabular}
\end{center}
\end{table*}

\begin{table*}[h!]
\caption{Range of RMSE assessing effect of [9,21] lag of days of the median communicable period}
\label{t21daysrmserange}
\begin{center}
\begin{tabular}{ llllllllllll }
  \hline
  &\multicolumn{11}{c}{Province} \\
  \cline{2-12}
   &\scriptsize Beijing &\scriptsize Shanghai &\scriptsize Jiangsu &\scriptsize Zhejiang &\scriptsize Anhui &\scriptsize Guangdong &\scriptsize Henan &\scriptsize Hunan &\scriptsize Chongqing &\scriptsize Sichuan &\scriptsize Jiangxi\\
  \hline
\footnotesize Maximum &6&7&11&17&10&17&15&18&8&9&11\\
\footnotesize Minimum &5&2&8&3&4&4&4&1&4&4&2\\
  \hline
\end{tabular}
\end{center}
\end{table*}

\begin{table*}[h!]
\caption{{MAPE assessing effect of the quarantine rate}}
\label{tlmape}
\begin{center}
\begin{tabular}{ llllllllllll }
  \hline
  &\multicolumn{11}{c}{Province} \\
  \cline{2-12}
  $l_n$ &\footnotesize Beijing &\footnotesize Shanghai &\footnotesize Jiangsu &\footnotesize Zhejiang &\footnotesize Anhui &\footnotesize Guangdong &\footnotesize Henan &\footnotesize Hunan &\footnotesize Chongqing &\footnotesize Sichuan &\footnotesize Jiangxi\\
  \hline
  0.4 &0.016&0.012&0.021&0.011&0.01&0.011&0.007&0.012&0.004&0.016&0.011\\
  0.5 &0.02&0.015&0.019&0.003&0.006&0.013&0.008&0.016&0.011&0.012&0.013\\
  0.6 &0.016&0.006&0.021&0.009&0.01&0.019&0.007&0.014&0.004&0.012&0.008\\
  \hline
\end{tabular}
\end{center}
\end{table*}

%\begin{table*}[h]
%\caption{{RMSE assessing effect of the quarantine rate}}
%\label{tlrmse}
%\begin{center}
%\begin{tabular}{ llllllllllll }
%  \hline
%  &\multicolumn{11}{c}{Province} \\
%  \cline{2-12}
%  l &\footnotesize Beijing &\footnotesize Shanghai &\footnotesize Jiangsu &\footnotesize Zhejiang &\footnotesize Anhui &\footnotesize Guangdong &\footnotesize Henan &\footnotesize Hunan &\footnotesize Chongqing &\footnotesize Sichuan &\footnotesize Jiangxi\\
%  \hline
%  0.4 &4&4&10&10&7&10&7&8&1&6&7\\
%  0.5 &6&5&9&4&4&13&8&18&5&5&11\\
%  0.6 &5&1&10&7&7&19&7&11&1&4&6\\
%  \hline
%\end{tabular}
%\end{center}
%\end{table*}

We use epidemic and migration data from January 24 through February 15, 2020, to forecast confirmed case numbers on February 16-18, 2020, for a network of human mobility including 11 nodes (provinces) in mainland China. 
The spread of COVID-19 during the period kept on growing and did not reach a steady stage. 
The epidemic data is available through the link \cite{chinadatawebsite}.
Data sets of the migration part are obtained from the Baidu Migration \cite{baidumigwebsite}. This website provides the group size of people  emanating from each province, and the percent of migration from one province to another province.

\begin{figure}[!h] 
	\includegraphics[width=0.5\textwidth]{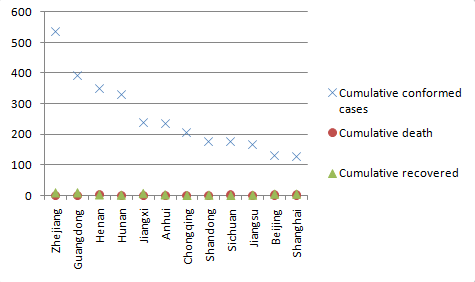}\\
	%\label{fig:1}
	\caption{{Cumulative numbers of diagnoses on January 30, 2020.}}
    \label{11province}
\end{figure}

We choose 11 Chinese provinces that were the most severely infected nationally during the time period, excluding Shandong and Hubei. The epidemic data of Shandong province are not included in this study as the cumulative number of confirmed cases dropped unusually on a period of consecutive three days.
Hubei province is not included in the analysis because of the Wuhan lockdown  that went into effect on January 24, 2020, cutting off the population mobility from those outside the province. 
Figure \ref{11province} displays the cumulative numbers of epidemic data such as confirmed cases, deceased cases, and recovered cases of the top 11 severely infected provinces aforementioned on January 30, 2020. The cumulative number of diagnoses in Shanghai is not available in the previously mentioned data link but can be obtained from the Shanghai Municipal Health Commission \cite{shanghaiwebsite}.  
%Using the migration data downloaded from January 24 to February 18, 2020, we could obtain the proportion values of $\gamma_n^{\textrm{in}}$, $\gamma_n^{\textrm{out}}$, $P_{nm}^{\textrm{in}}$, and $P_{mn}^{\textrm{out}}$. 

%Then we can apply the SaucIR model (\ref{SaucIR}) to predict the accumulated confirmed cases. 
In the following analysis, we first evaluate the effects of components of the model and the sensitivity of parameters involved in the model. We then assess the prediction fidelity of the proposed SaucIR model and compare the prediction result based on the epidemic and migration data in mainland China. 
%sensitivity of proportion of asymptomatic carriers and the length of median CP for the asymptomatic process. Next, we evaluate the the sensitivity of length of incubation period for the symptomatic process. Then we evaluate the the sensitivity of quarantine rate. 
%robustness of the model regarding 
%different selection of  parameter, based on pathological characteristics. 

%Firstly, we evaluate the sensitivity of parameters in the asymptomatic process, including the proportion of asymptomatic carriers and the length of median CP. 
%Secondly, we evaluate the sensitivity of length of incubation period in the symptomatic process. 
%Next, we use simulations to compare the effect of choosing different $l_n$.  
%Finally, we make simulation and prediction, and evaluate our model. 

\begin{table*}[h!]
\caption{MAPE evaluation of prediction of confirmed cases China nationally }
\label{t3}
\begin{center}
\begin{tabular}{llllllllllll}
  \hline
  &\multicolumn{11}{c}{Province} \\
  \cline{2-12}
  \footnotesize Method &\scriptsize Beijing &\scriptsize Shanghai &\scriptsize Jiangsu &\scriptsize Zhejiang &\scriptsize Anhui &\scriptsize Guangdong &\scriptsize Henan &\scriptsize Hunan &\scriptsize Chongqing &\scriptsize Sichuan &\scriptsize Jiangxi\\
  \hline
  \footnotesize SIR &0.036&0.018&0.015&0.007&0.026&0.035&0.016&0.017&0.014&0.038&0.041\\
  \footnotesize SIR+M & 0.039&0.024&0.012&0.007&0.026&0.036&0.015&0.016&0.014&0.038&0.041\\
  \footnotesize SaucIR-M & 0.018&0.006&0.023&0.011&0.016&0.021&0.007&0.021&0.019&0.015&0.016\\
  \footnotesize SaucIR & 0.017&0.006&0.015&0.011&0.006&0.015&0.005&0.014&0.009&0.009&0.010\\
  \hline
\end{tabular}
\end{center}
\end{table*}
\begin{table*}[h]
\caption{RMSE evaluation of prediction of confirmed cases China nationally}
\label{t4}
\begin{center}
\begin{tabular}{llllllllllll}
  \hline
  &\multicolumn{11}{c}{Province} \\
  \cline{2-12}
 \footnotesize Method &\scriptsize Beijing &\scriptsize Shanghai &\scriptsize Jiangsu &\scriptsize Zhejiang &\scriptsize Anhui &\scriptsize Guangdong &\scriptsize Henan &\scriptsize Hunan &\scriptsize Chongqing &\scriptsize Sichuan &\scriptsize Jiangxi\\
  \hline
  \footnotesize SIR &16&5&8&7&23&45&21&15&7&21&36\\
  \footnotesize SIR+M &16&7&7&7&23&47&21&14&7&21&36\\
  \footnotesize SaucIR-M & 6&2&15&11&13&24&9&20&12&8&13\\
  \footnotesize SaucIR & 5&2&8&10&5&17&7&13&5&5&8\\
  \hline
\end{tabular}
\end{center}
\end{table*}

Firstly, we evaluate the effects of the asymptomatic component in the SaucIR model, including the proportion of asymptomatic carriers and the  communicable period of asymptomatic carriers. 
%For the asymptomatic process, the sensitivity of proportion of asymptomatic carriers and the length of median CP need to be identified. 
%As the proportion of asymptomatic carriers is uncertain \cite{PFMS2020}, 
%For the proportion of asymptomatic carriers,
On one hand, we conduct simulations studying the effects of different levels of the rate of asymptomatic people ($\theta$) on the prediction accuracy. The results are summarized in Tables \ref{t1} and \ref{t2}.  The values of $\theta$ may be different spatially. Our simulation results assess and show that the SaucIR model works well when  $\theta$ takes values between 0.15 to 0.45 in terms of MAPE and RMSE. Both MAPE and RMSE show a similar pattern, such that the measure error first decreases and then increases as $\theta$ varies starting from 0.05 and upward to 0.45.  The bottom points of  $\theta$  of  measurement errors are slightly different among provinces. Thus, it may be concluded that $\theta$ affects the error of the predicted results but the prediction accuracy may be guaranteed when it is controlled under 0.45. Hence, we uniformly use 0.25 as the rate of the asymptomatic infected group to analyze the predictive fidelity of the proposed model.   Another observation is that the supercity Beijing has relatively larger values of MAPE and RMSE, which are highly likely due to the relatively extensive amount of imported cases based on the news and media report. 
On the other hand, we validate it is reasonable to take 21 days as the communicable period of asymptomatic carriers. 
%of time that {\color{red} [I AM NOT SURE WHICH CONJUNCTION IS APPROPRIATE, BUT ``that'' MIGHT NOT BE SOUND]} an asymptomatic infected person is contagious. 
We conduct a sensitivity analysis %use simulations 
to compare the effect of selecting different communicable period on the prediction accuracy. According to the literature \cite{hu2020clinical}, we select the length of time consisting of 6-21 days in the simulations. Both RMSE and MAPE reflect that the prediction error is smaller in most areas when the length of time varies between 9 and 21 (Tables \ref{t21daysmape} and \ref{tday21rmse}), and the effect of the value on the prediction error is small (Tables \ref{t21daysmaperange} and \ref{t21daysrmserange}).

%For the symptomatic process, the sensitivity of length of the incubation period need to be verified. 
Secondly, we evaluate the sensitivity of the incubation period in the symptomatic process. We conduct a sensitivity analysis to check whether the spread results change substantially for various incubation period. Considering the incubation period in the literature \cite{lauer2020incubation}, we take 3 to 7 days on the spread of the disease. Both RMSE and MAPE reflect that the incubation period corresponding to the best forecast varies from region to region. When the incubation period is between 3 and 5 days, the forecast error is small in most areas (Tables \ref{t5daysmape} and \ref{t5daysrmse}), and the effect of changing incubation period in this interval is small (Tables \ref{t5daysmaperange} and \ref{t5daysrmserange}). Consequently, 5 days is acceptable uniformly.

\begin{figure*}[!h] 
	\centering
	\includegraphics[width=0.9\textwidth]{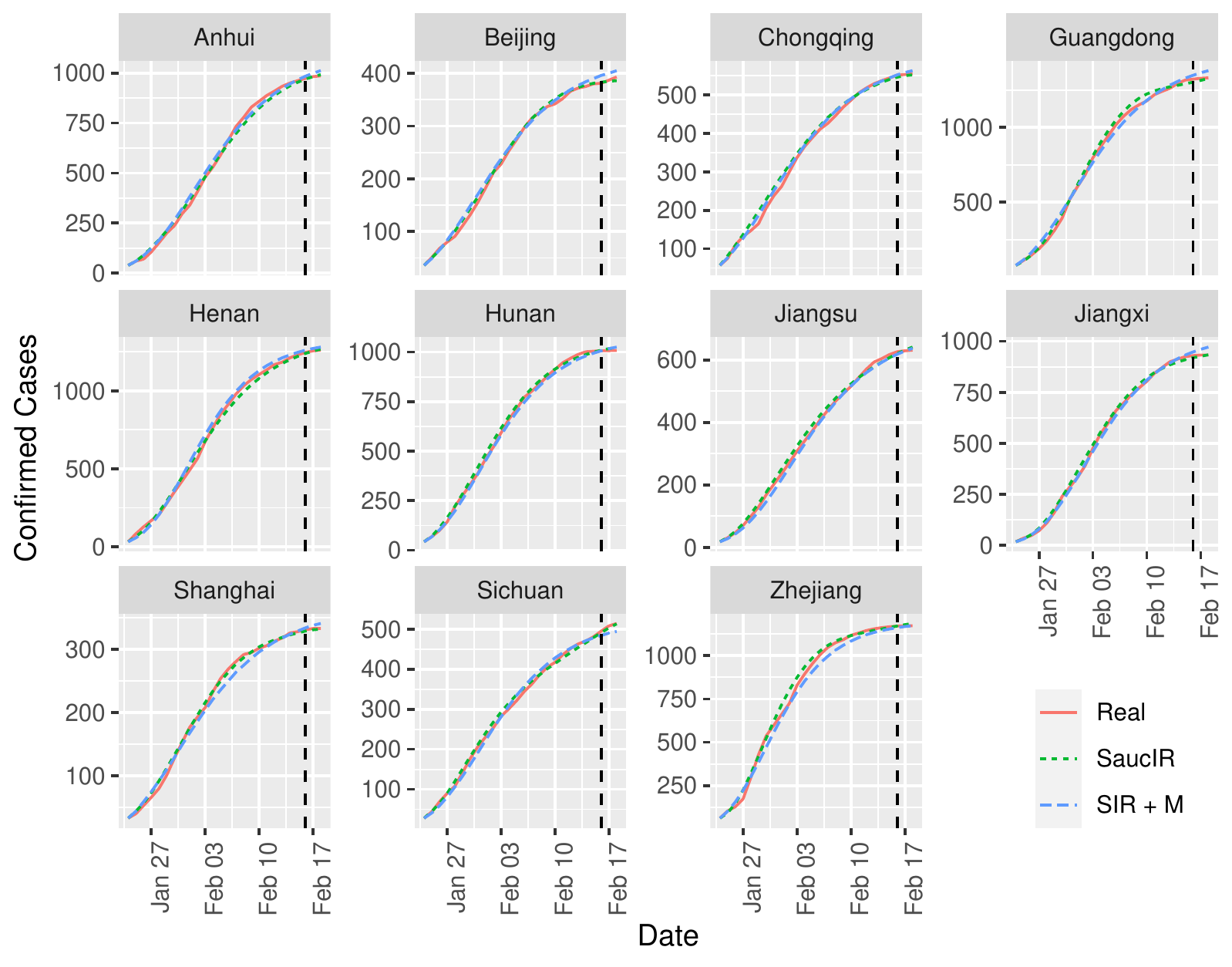}\\
	%\label{fig:1}
	\caption{Predicted numbers of confirmed cases in selected provinces. The data on January 24 through February 15, 2020 were used to fit curves while the fitted models were used to forecast the confirmed case numbers on February 16-18, 2020. }
    \label{provincePrediction}
\end{figure*}

%For the value of quarantine rate ($l_n$), 
Next, we use simulations to compare the effect of choosing different $l_n$. As mentioned in Section 2.3, it is reasonable to keep $l_n$ constant over time. Because there are no inflection points in Figure \ref{provincePrediction}, and the gains from making $l_n$ change over time come mainly from the inflection point. We then use simulations to compare the effect of choosing different $l_n$ fixed with time. Since the estimated $l_n$ varies between 0.4 and 0.6 in the real data, we consider three values of $l_n$ (i.e., 0.4, 0.5, and 0.6) for each node. Both MAPE and RMSE show that the model is insensitive to the value of $l_n$ (Table \ref{tlmape}). 

%Next, we explain why we do not make $l_n$ vary over time. 
%The previous method \cite{chowell2003sars} has $l_n$ changing over time, where the real data curve has a clear inflection point. However, our real data has no clear inflection point, so it is reasonable that varying $l_n$ over time will improve limited prediction accuracy. As observed in Figure \ref{quarantine}, large prediction accuracy gains due to varying $l_n$ over time occur only when the real data inflection point is obvious. 
%Moreover, as the sample size increases, varying $l_n$ over time results in larger gains in absolute terms.

\begin{table*}[h!] 
\caption{Predicted numbers of confirmed cases.}
\label{countryPredic}
\begin{center}
\begin{tabular}{ llllllll }
  \hline
  &\multicolumn{7}{c}{Country} \\
  \cline{2-8}
  Date & Italy &Spain &German &USA &France &South Korea &UK\\
  \hline
  05.19.2020 (Pred) & 225605&278216&177821&1548974&181245&10992&249210\\
  05.19.2020 (Obs)& 225886&278188&177289&1550294&180051&11078&247709\\
  05.19.2020 (PE)& -0.0012&0.0001&0.003&-0.0009&0.0066&-0.0078&0.0061\\
  05.20.2020 (Pred) & 226315&279774&178518&1564915&181651&10994&252171 \\
  05.20.2020 (Obs)& 226699&278803&178150&1570583&180934&11110&250141\\
  05.20.2020 (PE)& -0.0017&0.0035&0.0021&-0.0036&0.004&-0.0104&0.0081\\
  05.21.2020 (Pred) & 226997&281281&179194&1580095&182035&10996&255017 \\
  05.21.2020 (Obs)& 227364&279524&178748&1592723&181700&11122&252234\\
  05.21.2020 (PE)& -0.0016&0.0063&0.0025&-0.0079&0.0018&-0.0113&0.011\\
  \hline
\end{tabular}
\end{center}
Pred, predicted number; Obs, real number; PE, percentage error for every country. 
\end{table*}

Finally, we evaluate the prediction performance of our method through simulations. We demonstrate that the proposed segmentation of the infection compartment significantly improves the prediction accuracy and that the incorporation of dynamic human mobility apparently enhances the prediction accuracy. We use SIR+M to denote model (3) in the literature \cite{Brockmann1337}, and use SIR and SaucIR-M to denote models (\ref{SIR}) (without further segmentation) and (\ref{eqnInitial}) (without dynamic human mobility) in Section 2.1, respectively. The comparison results are summarized in Tables \ref{t3} and \ref{t4}. The predicted RMSE is in single digits except for Zhejiang, Guangdong, and Hunan. 

%Figure \ref{provincePrediction} compares the real and estimated confirmed cases during the period Jan. 24 to Feb. 15, 2020, and the forecast of the next three days. The real confirmed cases number from Jan. 24 to Feb. 15, 2020 were used to fit $\alpha_n$ and $\zeta_n$. The closeness between the two curves demonstrated a reasonable fit, validating the predictive fidelity of the SaucIR model that takes the epidemiological characteristics and human mobility into consideration.

Figure \ref{provincePrediction} shows the real numbers of confirmed cases and the fitted numbers of confirmed cases by SIR+M 
%{\color{red}the method of Brockmann and Helbing \cite{Brockmann1337} (abbreviated as the BH method)}
and our method during the period from January 24 to February 15, 2020, and forecast numbers of confirmed cases for the next three days. In both the SIR+M and our method, the real numbers of confirmed cases from January 24 to February 15, 2020 are used to estimate epidemic parameters. %$\alpha_n$ and $\zeta_n$. %The cross is real data. Green line is our method. Blue line is 's method. The left side of the dashed line is the fitted results and the right side is the predicted results. 
In most provinces, the forecast accumulated numbers of confirmed cases by SIR+M are generally larger than the real magnitudes. Our conjecture is, in SIR+M, the decrease of the transmission rate $\alpha$ is the main factor that affects the slowdown in the growth of the number of confirmed cases, whereas in SaucIR, the group sizes of the segments $U$ and $A$ will take effects simultaneously together with the transmission rate $\alpha$. Notice that, the partition of $I$ into $A$ and $U+C$ shown in Figure 1 implies that the size of the infectious compartment in SaucIR is smaller than that in SIR+M because the group of $C$ loses transmissibility owing to isolation for therapy. Also, the forecasting results by SaucIR are closer to the real numbers compared to those by SIR+M in most provinces because of the involvement of multiple factors of the epidemic and human mobility.

\subsection{Prediction in the international network of human mobility}

In this subsection, we use the international air transport network to demonstrate the predictive ability of the proposed SaucIR model. 
We use the accumulated confirmed cases from February 24 to May 18, 2020 to fit the epidemic parameters  and predict the confirmed cases from May 19 to May 21, 2020 for the seven severely infected countries. Confirmed cases and human mobility data of countries that ranked in the top six of confirmed cases and South Korea are analyzed. See page 4 of \cite{WHOwebsite}. Confirmed cases are announced by countries. The population mobility data are obtained by the airport announcements (c.f. \cite{oagmigwebsite}). As the quarantines of people (unconfirmed cases $U$ and asymptomatic carriers $A$) are not announced to the public, we set $l_n=0$. Then, we can calculate the number of confirmed cases in the future using model (\ref{SaucIR}).
 
%Table \ref{countryPredic} reports the forecasting confirmed cases during May 19-21 based on the previous nearly three-month epidemic and mobility data. Korea had a much lower amount of confirmed cases than other countries because instant prevention and monitoring interventions were implemented nationallingy.

%Figure \ref{countrydiffe} compares the curves of forecasting and real daily confirmed cases during the period Feb. 24 to May 21, where the red end in each subfigure indicates the forecasting results. Again, the closeness between the two curves demonstrates a reasonable fit, validating the predictive fidelity of the SaucIR model. One may also observe that the curve of the newly confirmed cases flattened almost from April, indicating the worldwide prevention measures, such as reducing the population mobility strategy by cutting off the intercountry air mobility, are taking effect.

\begin{figure*}[!h] 
	\centering
	\includegraphics[width=0.9\textwidth]{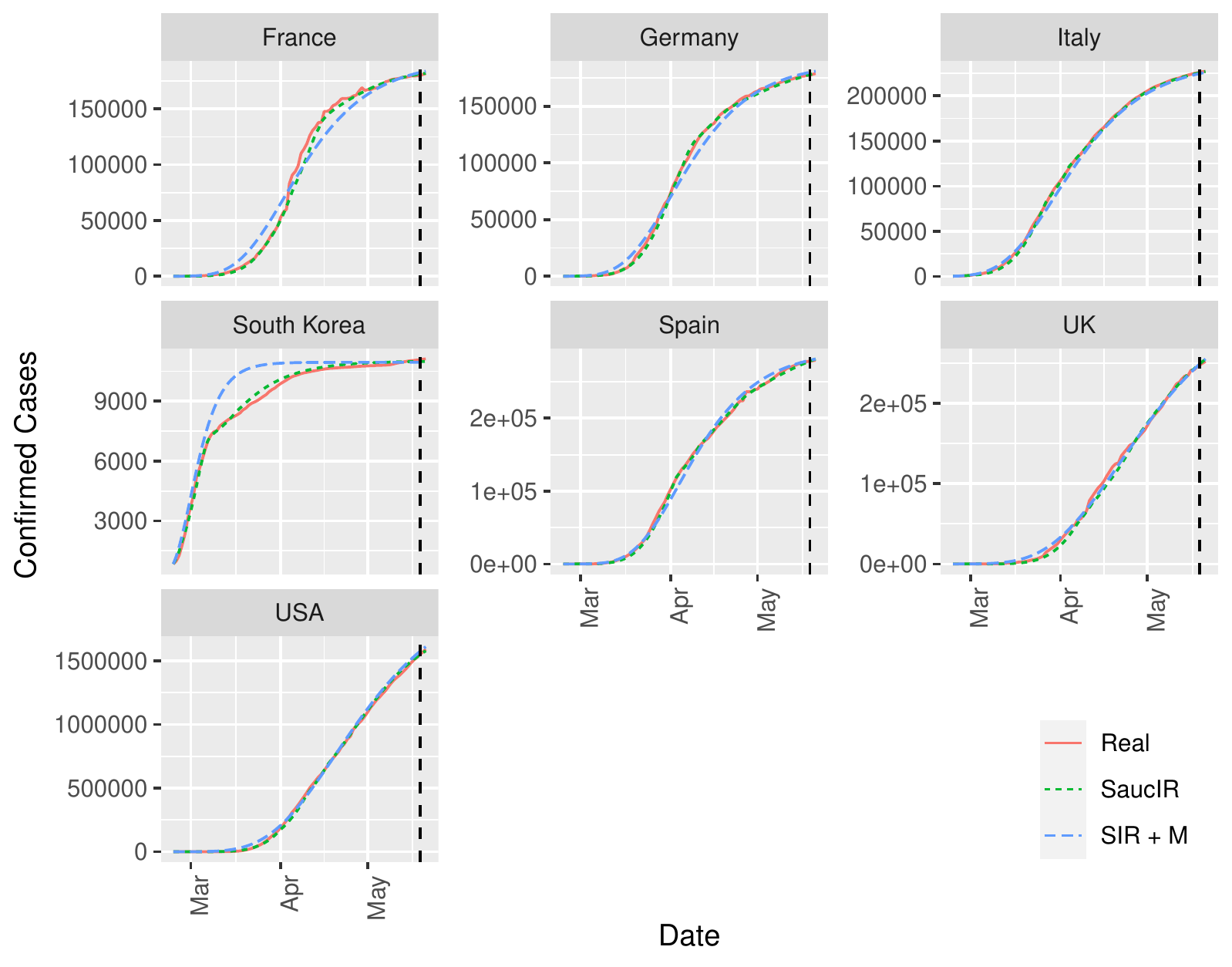}\\
	%\label{fig:2}
	\caption{Predicted numbers of confirmed cases in selected countries. The data on February 24 through May 18, 2020 were used to fit curves while the fitted models were used to forecast the confirmed case numbers on May 19-21, 2020. }
    \label{countrydiffe}
\end{figure*}

Table \ref{countryPredic} compares the predicted results and the real data during May 19-21. Evidently, our prediction has smaller relative prediction errors.
Figure \ref{countrydiffe} graphically displays the predicted curves by our method and the SIR+M together with the observed values for seven countries during the period from February 24 to May 21. 
%The part before April 30 has been compressed to show the prediction part better. 
%is selected to compare our method,  method and the real data. 
In those countries where SIR+M fits well, the fitting and forecasting are well consistent with the characteristics we mentioned in Section \ref{sec3.1}. For those countries where SIR+M does not fit well (e.g., France), the reason could be that not all infected persons can be tested immediately \cite{germanycovid} so that the cumulative number of confirmed cases and the number of infected people are not exactly the same. Our method is more capable to reflect this due to the segmentation of $U+C$. Therefore, our method has forecasting results closer to the real data, with the maximal MAPE of 0.011.

\subsection{Impact of migration on confirmed cases}
In this subsection, we study how the migration scales affect the confirmed cases. The epidemiological parameters and mobility parameters involved in the dynamic SaucIR model (\ref{SaucIR}) are calculated or fitted based on the migration data obtained from the Baidu website and epidemic data announced officially from Janruary 24 to February 29, 2020. 

%As mentioned in Section 2.4, we calculated the series of $(\gamma P)_{nm,t}^{\textrm{in}}$ and $(\gamma P)_{nm,t}^{\textrm{out}}$. 
The migration scale ($C_{nm,T}^{\textrm{in}}$ and $C_{nm,T}^{\textrm{out}}$) is categorized into three levels, where the large and medium scales are triple and twice of the small scale. Using optimal procedure (\ref{optimal}), the coverage of confirmed cases can be obtained under various migration scales.

Table \ref{Impactofmigration} summarizes the predicted confirmed cases on the nodes in the 11-province human mobility network of mainland China. Vertically, in the same scope of migration scale, the estimates confirmed cases do not vary much between the minimum and maximum. This implies that modifying the emanating  rates from a node in or out of another node within the network has a minor impact on controlling the transmission rate. Horizontally, a smaller migration scale, less confirmed cases. This may be evidence of the validity of Wuhan lockdown for preventing the spread of COVID-19.

\begin{table}[H] 
\caption{Influence of migration scales on confirmed cases}
\label{Impactofmigration}
\begin{center}
\begin{tabular}{ llll }
  \hline
   &\multicolumn{3}{c}{$C_{ij}$} \\
  \cline{2-4}
  Confirmed cases & Large &Medium &Small \\
  \hline
  Maximum & 9481 & 9377 & 9268  \\
  Minimum & 9439 & 9344 & 9251  \\
  \hline
\end{tabular}
\end{center}
\end{table}

%\FloatBarrier
\section{Conclusion and Discussion}

Compared to the SIR model raised in \cite{Brockmann1337}, our SaucIR model is disease specific for the spread of the COVID-19 pandemic. The primary difference is that the infection compartment has an independent segment of asymptomatic infected individuals based on the epidemiological characteristics of COVID-19. The numerical results indicate that such re-compartment is critical and substantial. By the time our revised manuscript was submitted, we found more studies on asymptomatic with considerable proportions up to 58\% among patients with infection, justifying the rationale of independent A segment in the proposed compartmental model. The rate of asymptomatic per sons in population-based retrospective studies can even reach 43\% \cite{DF2020NewEJM}. It remains a series of unsolved problems regarding the asymptomatic patients in COVID-19 that are anticipating more epidemiological observations and experiments, such as pattern of infectiveness, accurate communicable period,  effective surveillance and screening \cite{leejamainternmed2020}. The next distinction is that we present a dynamic migration procedure rather than fixing the mobility parameters. Baidu migration data source makes it feasible for modeling and computing. The advantage of such dynamic human mobility makes it possible to take into consideration of infection during the procedures of migration in and out of a spatial node, achieving the predictive fidelity of the proposed SaucIR model. 
On the other hand, it validates the necessity of the separation of moving in and out of the same node from other nodes in the modeling.  The minor difference is that we use the absolute flux of population rather than the fraction of the passenger flux. The big population size of the provinces in China brings out a fraction of each compartment far less than unity, which has impact on the prediction accuracy. We need to point out that our modeling uses newly published parameters such as incubation days of infected to the symptom and asymptomatic to recovered. Modeling the spread pattern is an important issue and 
could be updated when there are new and reliable epidemic parameters.

The COVID-19 pandemic has not ended yet. The forecasting SaucIR model will benefit the public health authorities to assess the epidemic situation, make reasonably informed decisions,  take appropriate interventions, and give a timely control of infection.

\section*{Acknowledgements}
The authors thank the invitation from the co-chief editors and two reviewers for constructive comments. The authors also thank Dr. Sheng Xu for assistance in editing of figures.

%\input{FCCCbyMNov.bbl}
%\bibliographystyle{imsart-number}  %{plain}
%\bibliography{FCCCbyMNov}

\end{document}